\documentclass[aps,preprint, superscriptaddress]{revtex4-1}

\usepackage{color}
\usepackage{graphicx}
\usepackage{epstopdf}
\usepackage{color}

\begin{document}

\title{Microparticle transport networks with holographic optical tweezers and cavitation bubbles}

\author{P. A. Quinto-Su}
\email{e-mail: pedro.quinto@nucleares.unam.mx}
\affiliation{Instituto de Ciencias Nucleares, Universidad Nacional Aut\'onoma de M\'exico, Apartado Postal 70-543, 04510, Cd. Mx., M\'exico.}






\begin{abstract}
Optical transport networks for active absorbing microparticles are made with holographic optical tweezers. The particles are powered by the optical potentials that make the network and transport themselves via random vapor propelled hops to different traps without the requirement for external forces or microfabricated barriers. The geometries explored for the optical traps are square lattices, circular arrays and random arrays. The degree distribution for the connections or possible paths between the traps are localized like in the case of random networks. 
The commute times to travel across $n$ different traps scale as $n^2$, in agreement with random walks on connected networks. Once a particle travels the network, others are attracted as a result of the vapor explosions. 
\end{abstract}



\maketitle

Developing autonomous machines at the micro and nano spatial scales requires engines \cite{berg, kinesin1, kinesin2, ma} and
the ability to control transport to targeted spatial locations \cite{koumakis1}.  The most promising proposals include active particles \cite{bechinger} that convert energy from the environment into motion. So far, targeted transport has only been achieved using biological active particles in combination with microfabricated substrates, barriers and channels \cite{dileonardo1, mahmud, volpe1, reichhardt}. 

Brownian active particles are systems that are out of equilibrium due to the absorption and conversion of energy from the environment. Some particles are heated by light, and motion is the result of an induced local temperature gradient (self thermoforesis) like in the case of Janus particles \cite{janus1} that have an asymmetric absorption coefficient. Other particles can catalize chemical reactions \cite{phoretic} for self propulsion like platinum-copper nanorods immersed in dilute Br$_2$ or I$_2$ solutions (self-electrophoresis) \cite{nanorods}. Living bacteria have also been used as active particles or as active baths that interact with artificial microbeads \cite{argun}.

The motion of active particles can be described by anomalous diffusion which can be controlled by the energy in the environment \cite{bechinger}. 
In several applications these particles have to be physically constrained so that they do not leave the area of interest, for example: swiming bacteria in microfabricated environments like gears and walls \cite{dileonardo1, volpe1}. 

Regular Brownian microbeads can also be driven out of equilibrium to induce directed motion when combined with potentials and external forces, making Brownian ratchets \cite{facheux, nanoratchet, arzola}.
The main limitation of these ratchet systems is that many parameters have to be carefully tuned to enable directed transport: the potentials have to be periodic (usually spatially asymmetric) and external forces are required in most implementations. 

\begin{figure}
 \includegraphics[width=3.3in]{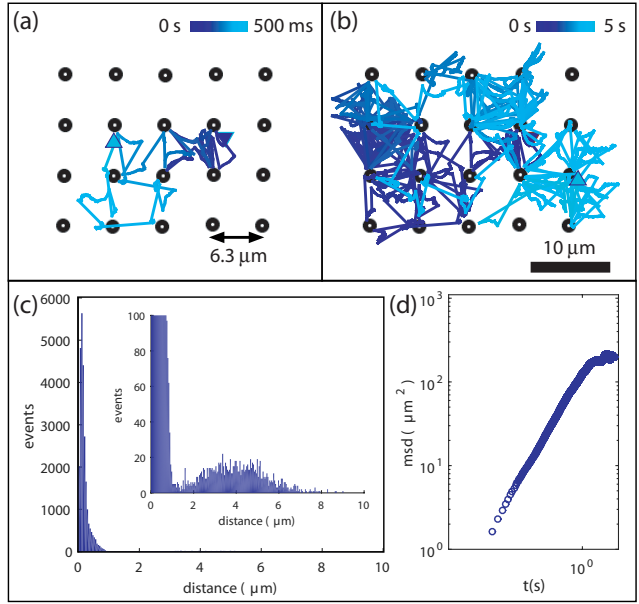} 
 \caption{(a) Microparticle trajectory spaning 500 ms. (b) Trayectories spanning 5 s. Particle initial position: dark blue triangle, final position light blue triangle. The color gradient represents time (dark to light). The circles are the position of the optical traps (waists of the focused laser beams). (c) Step distribution. Inset: zoomed in step distribution. (d) Mean squared displacement (msd) for the particle. Data in (c-d) extracted from 27000 frames (27 s) with 1200 events (hops). The frequency of the jumps is $52 \pm 22$ Hz. Video  1.}
\end{figure}
%

In this study we use an absorbing microbead (magnetic) that is attracted to the waist of a focused laser beam superheating a small volume of liquid creating an explosion or a cavitation bubble (lifetime of a few microseconds) that ejects the particle from the trap \cite{steam1}. Hence, the effective potential is similar to a Lennard-Jones potential that is attractive at long distances and repulsive at short range. In an optical potential landscape created with highly focused laser beams, the particle propels  itself and becomes a random walker hoping between the different traps that effectively make a transport network. 
There is no need for external forces or the tight constraints that regular ratchet systems require. The particles travel the two dimensional network and do not leave the area of interest like in the case of other active particles \cite{dileonardo1, volpe1, janus1, nanorods}. 

The experiment is done in a standard holographic optical tweezers setup with a 2.5W, 1064 nm trapping laser.  The optical potential landscape is generated with a spatial light modulator (SLM, Hamamatsu X10468-07) and the digital holograms are calculated with a Gershberg Saxon algorithm \cite{gs} to generate arrays of focused laser beams.
The microbeads are magnetic (Bangs Promag) with a mean diameter of 3.16$~\mu$m immersed in water. The events are imaged by a high speed camera and are recorded at 1000 frames per second (fps) with a frame size of 256x320 px, limiting the recording time to 68 s.

Figure 1a-b shows the trajectory of a single microparticle in the xy plane in an array of 20 optical traps (black circles) making a square lattice (See Video  1). 
The trajectory starts at the darkest triangle and the color gets lighter with time (500 ms in Fig. 1a and 5s in Fig. 1b).
The trajectories resemble a truncated Levy walk \cite{levy1} were most of the steps are smaller than $1 ~\mu$m (Fig. 1c) and are directed to a potential well, while a smaller fraction of the steps are the large vapor propelled hops. The inset of Figure 1c shows the zoomed step distribution that is long tailed. We observe that the vapor propelled hops have a range between 1 and 10 $\mu$m with center at about 4 $\mu$m. 

Interestingly, the particle is constrained to move only in the area spanned by the array of traps, so the mean squared displacement (msd) in Fig. 1d resembles that of a regular Brownian particle trapped with optical tweezers \cite{volpe2}, where the msd increases monotonically until it reaches a steady state value that characterizes an effective area where the particle is moving. 
This result contrasts with what is typical of active particles where the msd increases without bound \cite{bechinger}, making the particles leave the area of interest if not constrained by physical barriers.
In the case of particles trapped in regular optical tweezers the msd limit is of a few hundred squared nanometers \cite{volpe2}, while in these experiments that value exceeds one hundred squared micrometers (Fig. 1d).

\begin{figure}
\includegraphics[width=3.3in]{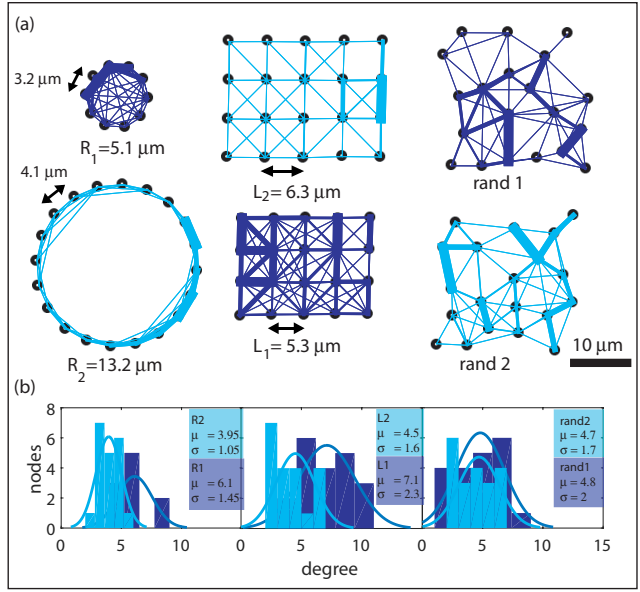} 
\caption{Microparticle transport networks. (a) Graphs representing the transport networks. Videos 1-6. (b) Node degree distribution for the different networks. }
\end{figure}


The graphs for the transport networks that we explore are shown in Fig. 2a: circles with 10 traps (radius $R_1=5.1 ~\mu$m) and 20 traps ($R_2=13.2~\mu$m). Square lattices with 20 traps ($L_1 =5.3 ~\mu$m and $L_2 =6.3 ~\mu$m). Randomly placed traps, rand 1 and rand 2 have minimum separations of $4.1-4.2 ~\mu$m and maximum separations of $6.8-6.3 ~\mu$m respectively.

We track the trajectories of a single particle across the two dimensional arrays of optical potentials (Videos 1-6) in the different graphs. The traps that are connected by the particle trajectory are joined by lines. 

For an array of $n$ traps (nodes) the $n\times n$ possible connections between all the nodes can be represented by an $A_{ij}$ array called the adjacency matrix. Typically the entries $A_{ij}$ in the adjacency matrix have a value of 1 if the $ij$ nodes are connected and 0 if those are disconnected. 

In our experiments the networks are weighted, which means that there are some paths that are more transited for the different $ij$ pairs. This property is characterized by the weighted adjacency matrix $w$, where the entries $w_{ij}$ are the number of transits between traps $i$ and $j$. In this way the total number of transits is $T=\sum_{ij} ^n w_{ij}$. 
The widths of the lines in Fig. 2a are proportional to $(w_{ij}+w_{ji})/2$.

The transitions $w_{ii}$ that start and end at the same node are not drawn (Fig. 2a). The total number of $w_{ii}$ transits is $T_s=\sum _i ^n \ w_{ii}$ and the fraction of these transitions is $T_s/T$. These self connections are more common when the separation between the nodes increases and for nodes at the perimeter of the graphs because those nodes are less connected. The number of self connections vary between $8\%$ of the total transits for the small circular graph to $43 \%$ for the second random graph (rand 2).

The number of connections for a given node $j$ is described by the 
degree ($k_j$):
\begin{equation}
k_j=\sum _{i}^{n} A_{ij}.
\end{equation}

Figure 2b depicts the degree distributions, where the vertical axis is number of nodes that have a given degree. 
We observe that in all cases most of the nodes have a similar number of connections resulting in a localized distribution. This is expected for random networks, which can describe some transportation networks like highways \cite{barabasi}.  
The histograms in Fig. 2b are fitted to Gaussians that yield a mean degree and a standard deviation. 

\begin{figure}
\includegraphics[width=3.3in]{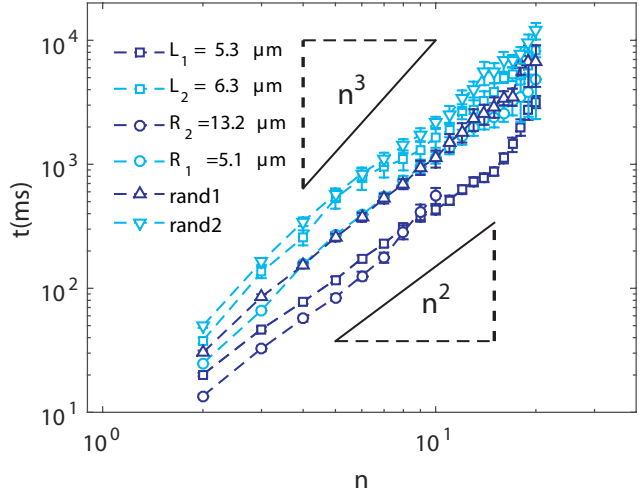} 
\caption{Commute times to travel across $n$ different nodes on the networks. Data for the cases depicted in Fig. 2. The symbols represent the average time to travel across $n$ different traps and the error is the standard deviation divided by the square root of the number of samples. The round symbols are for the circular graphs, squared for the squared lattices and the triangles for the random graphs. }
\end{figure}


The circle with the largest diameter $R_2=13.2~\mu$m has a mean degree of $3.95$ that shows that most of the traps are connected to the two contiguous traps (2 nearest to the left and right). Most of the traps in the circular array with the smallest radius $R_1=5.1~\mu$m  are connected to six others which are more than half of the total of 10 traps. 
The square lattice with inter trap separation of $L_2= 6.3~\mu$m has essentially only nearest neighbor connections (up, down, diagonals, mean degree of $4.5$), while the smaller separation $L_1 = 5.3~\mu$m enables connections to traps that are farther away (mean degree of 7.1). 
The random graphs have similar degrees ($4.8$ and $4.7$ for rand 1 and rand 2 respectively).


The trajectory of each particle across the graphs yields a sequence of the traps that are visited and the arrival times to each location. In this way we can measure the time it takes the particle to visit different traps. 
The time to travel from node $i$ to node $j$ is the commute time. Here we measure the time it takes the particle to travel across $n$ different nodes. This is done by finding the sequences  that contain $n$ different traps in the time series of visited traps and arrival times. As $n$ increases the number of samples decreases because the time it takes to reach a rising  number of different nodes increases monotonically. 
Finally, we average the extracted commute times to travel $n$ different traps and the results are plotted in Fig. 3.

We observe (Fig. 3) that in the case of graphs that have the same geometry ($R_1 -R_2$ and $L_1 -L_2$) there is an inverse relation between average commute time and mean degree as expected, because increasing the number of paths available to other nodes increases the speed of transport, decreasing the commute time.

The time it takes the particle to reach all the nodes in a graph is the cover time. Both, the commute time and cover time for random walks on connected graphs have an absolute upper bound $\propto n^3$ \cite{lovasz, feige}. In the case of a regular graph where each vertex is connected to the same number of neighbors, the cover time and commute times are proportional to $n^2$ \cite{lovasz}.

The measured commute times are proportional to $n^2$ (Fig. 3) and are bounded by $n^3$ in agreement with the theory for random walks on connected graphs. 
For Brownian motion, the square of the distance traveled is proportional to elapsed time. In this case the distance traveled is also proportional to the number of different nodes $n$. Hence, the average commute time to travel across $n$ different nodes is proportional $n^2 $.

Also, the speed of transport is proportional to the hopping frequency, which will depend on the initial displacement by the particle during the explosion and on the distance between the traps. In the case of the circular graphs the frequencies are $103 \pm 54$ Hz and $75 \pm 49$ Hz ($R_1$ and $R_2$),  $52\pm 22$ Hz and $76 \pm 44 $ Hz for the square graphs ($L_2$ and $L_1$), $65\pm 45$ Hz and $49 \pm 37$ Hz for the random graphs (rand 1 and rand 2).  

\begin{figure}
\includegraphics[width=3.3in]{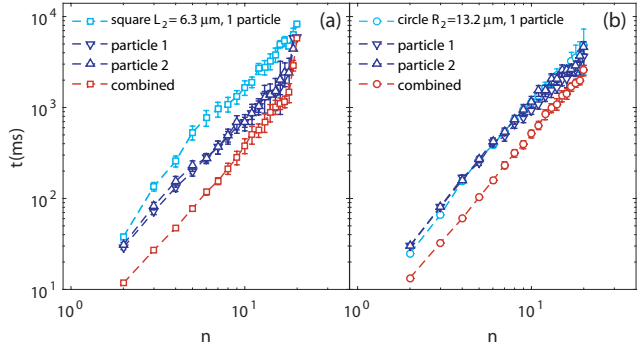} 
\caption{(a) Commute time comparison between one particle and two particles in square lattice $L_2=6.3~\mu$m. (b) Commute time comparison between one particle and two particles on the large circular array $R_2=13.2~\mu$m . Videos 8-9.}
\end{figure}


One property of these active particles or microscopic steam engines is that every time the particle hops, the vapor explosion also delivers an impulsive force \cite{partburb} to neighboring objects directed to the source of the explosion. Once a microparticle is traveling the network, the typical waiting time for other particles to eventually reach that area is on the order of tens of seconds ($\sim 30$ s for our dilute microparticle sample).
The particles that travel in the array of traps interact with each other via the vapor explosions, resulting in particles chasing each other (Video  7). However, if the particles converge in a trap it is likely that at least one es ejected due to larger vapor explosions \cite{twopart}.   

Figure 4a-b (Videos 8-9) shows the average commute times for two particles simultaneously traveling the two less connected networks: the square array with the largest separation ($L_2=6.3~\mu$m) and the large circle ($R_2=13.2~\mu$m). In the case of the square array (Fig. 4a) we observe that the interaction between the two particles enhances transport for the individual trajectories (dark blue  symbols) compared with the case of just one particle (light blue symbols); especially for transits that cover more than half of the nodes ($n \leq 16$). Once the particles get close to each other, the interactions slow transport until a large explosion ejects the particles at sufficient distance where the impulsive interactions  can enhance transport again. 
In the case of the large circle  $R_2$  (Fig. 4b), the wide radius makes it similar to a straight line. The impulsive forces have little effect because the surface density of traps is small, decreasing the number of traps where the particles can travel when their trajectory is perturbed by the neighboring vapor explosions. 
The red symbols in Figs. 4a-b represent the average commute times when considering both particles combined. This is done by sorting (by arrival times) and joining the time series (visited nodes, arrival times) of both particles. We observe a much larger enhancement, compared with the case of a single particle (light blue symbols) in the network, specially in the case of the square lattice when $n  \leq 16$.

To conclude, we have demonstrated active particle transport that reaches targeted spatial locations without the need of microfabricated barriers or external forces. The main mechanism is the effective potential that is attractive/repulsive at long/close range. The particles are confined in an arbitrary optical potential landscape made with holographic optical tweezers.
The time it takes a particle to visit $n$ different nodes scales as $n^2$. 
Once an active particle is traveling the network, the explosions attract other particles into the array of traps, enhancing the rate at which the nodes are visited and in turn pulling more particles into the network (Video  10). In this way, it is possible to have particles traveling the network for an arbitrary amount of time even if some are ejected, making the system robust.  
This property could be important for future micro/nano machines, as it can enhance the number of particles reaching an area of interest after one arrives there.



\section*{Funding Information}
Work partially funded by DGAPA UNAM PAPIIT grant IN104415, CONACYT CB-253706  and LN-299057. 
Thanks to Jos\'e Rangel Guti\'errez for machining several optomecanical components.

\bibliographystyle{apsrev4-1}



\end{document}